\documentclass[twocolumn,preprintnumbers,amsmath,amssymb]{revtex4}

\usepackage{graphicx}
\usepackage{dcolumn}
\usepackage{bm}
\usepackage{epsfig}
\usepackage{amssymb}
\begin{document}

\preprint{}

\title{Additional Comments on ``New Limits on Spin-Independent Couplings of Low-Mass WIMP 
Dark Matter with a Germanium Detector at a Threshold of 200 eV''}

\author{F.T. Avignone III$^{a}$, P.S. Barbeau$^{b}$ and J.I. 
Collar$^{b,*}$
}
\address{ 
$^{a}$Department of Physics and Astronomy, University of South Carolina, Columbia, SC 29208\\
$^{b}$Kavli Institute for Cosmological Physics and Enrico 
Fermi Institute, University of Chicago, Chicago, IL 60637\\
$^{*}${\frenchspacing Corresponding author. E-mail: collar@uchicago.edu}\\
}


\maketitle

In a previous comment \cite{us} we criticized purely technical aspects of 
\cite{texono} but pointed out that many other inconsistencies were readily apparent. 
In view of the imminent posting of another version of \cite{texono}, 
still containing these issues, we would 
like to offer with this additional note a critical guide to the 
methodology followed by the TEXONO collaboration, in their effort to extract dark 
matter limits from a spectral region completely dominated by electronic noise. 
We find that systematics have been neglected through a 
combination of erroneous assumptions and failure to explore all of their 
possible sources. 

We should start  by briefly putting the TEXONO method and results in 
perspective (in what follows all calls to figures or tables are for those 
in \cite{texono} unless 
otherwise stated). The raw counting rate in their ULEGe detectors prior to any 
data treatment is approximately $2\times10^{7}$, $7\times10^{5}$ and 
$3.5\times10^{3}$ 
counts / keV kg day in the bins spanning 0.1-0.2, 0.2-0.3 and 0.3-0.4 keV 
of ionization energy, respectively \cite{henry}. 
In the region entirely clear of the electronic 
noise (0.4-1 keV) this raw rate plateaus to $\sim10^{3}$ counts / keV kg day. 
Following a method of pulse shape 
discrimination (PSD) delineated further down, regions below threshold 
undergo a large stripping of what TEXONO claims is reliably mostly 
noise signals, supposedly preserving a significant fraction of true events. 
The bulk of this reduction is due to PSD and not 
vetoing from active elements in the shielding. The TEXONO collaboration then 
arbitrarily selects a 43 
eV-wide energy region (0.198-0.241 keV) for their light 
WIMP analysis, where the original $\sim$100,000 raw
events have been reduced to exactly zero by the effect of cumulative 
cuts (Table I). 
It is important to remark that similarly-sized neighboring regions contain a significantly larger number of 
surviving counts (e.g., five counts in the 0.241-0.300 keV region), 
i.e., TEXONO has selected this extremely narrow energy region in 
order to obtain 
the best possible light-WIMP sensitivity (see discussion below). TEXONO maintains that the 
error bars in the residual spectrum 
are eminently statistical (of order 50\%), dominating over systematic 
components.

We list below some of the systematic effects that have been 
neglected, and some of the assumptions that have been erroneously 
made. This list is by no means complete, for a close study of 
\cite{texono} 
reveals many open questions. In the interest of keeping this 
note in the realm of the readable, we limit ourselves to the most 
relevant issues only. 

$\bullet$ The language used in \cite{texono} would lead the unsuspecting reader to 
believe that ``software was devised to differentiate physics events from 
those due to electronic noise'', leading to ``the noise events being 
(were) suppressed''. The PSD method chosen to reject anomalous pulses 
has been extensively used 
in previous dark matter and low-background experiments \cite{morales,jcap,moralesother,cogent} with the clear 
difference that no collaboration before has dared to apply it well into 
the electronic noise pedestal. That such prudence is advisable 
should become clear in what follows.
The PSD method in question consists of comparing preamplifier 
signals routed through two shaping amplifiers with 
different time constants. Anomalous pulses such as microphonics 
(independently confirmed as such in \cite{morales})
are observed to exhibit different amplitudes (assigned energies) in these two channels, 
leading to their elimination. TEXONO offers no references to any previous 
use of this 
method. They have nevertheless unnecessarily adulterated it by obtaining energy 
information from the two channels using a slightly 
different approach for each. That these are ``partial'' and ``total 
integration'' is the sum of the details offered in \cite{texono}. 
Under our scrutiny, this adds no benefit to the method: 
much to the contrary, in private communications 
\cite{henry} we have learned that through non-linearities it may lead to a disagreement in the 
energy calibration between the two channels at the crucial 
low-energies. If confirmed this 
would manifestly defeat the validity of the method. Unfortunately such 
issues are veiled 
in Fig. 2b by 
choosing to represent the energy scale in one channel in keV and the 
other in arbitrary units. While we appeal to the TEXONO 
collaboration to be much clearer and extensive about critical points like 
this one, 
the bone of contention is a different one: contrary to what is stated 
in \cite{texono} the method described above is not capable of 
differentiating a ``physics'' signal from an electronic noise event 
at very low energies, well into the threshold. Both look the exact 
same. 
The reader 
should keep in mind that this is not the same situation as in other dark 
matter experiments, where calibrations allow to clearly distinguish 
desired recoil signals from minimum ionizing backgrounds, based on PSD. We invite the reader to 
inspect the discussion around Fig. 7 in \cite{jcap}, the first description 
of the method in \cite{morales}, and its 
conventional use in \cite{moralesother,cogent}. 
Even in such a dire situation, it may still be possible 
to apply an {\it entirely arbitrary} set of PSD cuts 
based on this method, as long as signal acceptance losses can be 
calculated and a correction to recover from those can be applied. This is the approach followed 
by TEXONO, in principle acceptable if it could be 
properly implemented. However, we strongly emphasize that the  
intrinsic arbitrariness in the placement of these cuts (see Fig. 2b) is 
in itself a source of systematic effects and errors, and therefore 
worthy of an investigation. This is absent in \cite{texono}.

$\bullet$ The TEXONO collaboration utilizes two methods to calculate signal 
acceptance losses from PSD cuts. In the first they expose their 
ULEGEs to several low-energy x-ray sources. No attempt is made to 
extract the spectral response to the sources from a comparison to 
background runs. Instead it is simply {\it assumed} that the response should be a 
flat spectrum below threshold ($<$ 0.3 keV). A comparison between this assumed 
spectrum and the spectral shape of data passing PSD cuts (Fig. 2a) is used to 
contribute to the correction curve for signal losses (black squares in Fig. 3). 
The lightheartedness with 
which this assumption is made is regrettable, for it negates a very 
large body of work in the 
area of x-ray detector response to low-energy photons. This 
response is a complex function of atomic and electron physics, 
detector structure and operating conditions, electronic contact 
design, window material, etc., and generally results into what is
anything but a flat distribution 
for sub-keV energies \cite{xray}. Several processes leading to partial energy 
deposition or charge collection are readily evident in the 
drastically raised 
plateau below the highest calibration peak in Fig. 2a, even 
if the spectrum is cut off immediately above in the figure.

$\bullet$ This first method of calculation of signal acceptance loss is clearly 
invalid in its present form. A very complex and dedicated effort would be 
necessary to extract more reliable information about the true low-energy response to the 
x-ray calibration sources, as indicated in the few example references in \cite{xray}. 
TEXONO exploits a second approach, which is to utilize low-energy events 
coincident with the vetoes as bona fide radiation-induced events. 
While this is a well-known technique \cite{jcap},
these coincident events are, we insist, in no other way distinguishable from 
electronic noise. This rises the question of what (energy-dependent) fraction of these 
are spurious noise coincidences with the vetoes, again a  
source of systematic errors. Others could be cited at this point: for 
instance, given the different response of the detector to events 
happening in or near the contacts \cite{xray} (here most of the detector surface)  such as external x-rays and 
betas, only sufficiently delayed coincidences should be used to insure a sampling 
with recoil-inducing neutrons. There is no indication that this has been taken into 
account. However, the most important fact is that this second method 
generates on its own a large uncertainty in the signal acceptance 
curve (``ACV tag'' in Fig. 3). This larger systematic impacts by itself the dark matter sensitivity 
and should generate relaxed limits.

$\bullet$ We are intrigued by the peculiar way in which the 
 quenching factor 
(QF) and its uncertainty
have been evaluated. 
Rather than relying on the good-quality experimental data 
available for 
the relevant low-energy region (see Fig. 4 in \cite{jcap}), TEXONO 
has used an outdated version of a simulation code, assessing experimental data out to 
200 keV to, in some way, estimate an uncertainty in its output at 
200 eV. That the Lindhard theory for the QF in germanium diverges at 
low and high recoil energies is clearly exposed in \cite{jones}, a 
reference TEXONO cites. We also notice 
how the most recent low-energy data, 
those in \cite{jcap}, have not been considered. 
All low-energy germanium QF measurements (four 
experiments, spanning 0.25-5 keV in recoil energy) are in excellent 
agreement. The route of fitting these relevant data 
instead would lead to a well-defined non-arbitrary 
uncertainty for the QF.
This more natural assessment of the QF can have a large effect in the 
claimed light-WIMP sensitivity. TEXONO should be cautious 
on the subject of this very important source of systematic error. 
A measurement of their own would be 
welcome.

$\bullet$ No method used to extract dark matter limits has ever gained universal 
acceptance. Fits using background models, no matter 
how reasonable, can be criticized for making some assumptions but 
then again 
lauded for employing all of the information in the region of 
interest, minimizing the beneficial rainfall from a favorable 
statistical fluke in some spectral region. 
Methods such as the unbinned optimal interval utilized by TEXONO 
have encountered the exact opposite commentary, conservative on one 
front 
and daring on the other. There is however a 
degree of 
audacity involved in its use on a spectral region just 43 eV wide, surrounded 
by wildly fluctuating concentrations of residual events, following a 
complete stripping of the raw data by five orders of magnitude, based 
on arbitrary cuts. For one thing, the 
degree of gain stability over time required for this may well be insufficient 
(``less than 5\%'' is mentioned, but unfortunately no reference energy 
is provided). At a minimum, in this extreme case a study 
of sensitivity as a function of chosen analysis region seems in order. A further 
comment is indicated, which is that rarely a dark matter search 
attempts
an analysis after an exposure of just 0.34 kg-days, 
precisely to reinforce the robustness of any method chosen to 
extract limits.

By daring to go where all other dark matter experiments have balked, 
TEXONO should expect a very high level of 
scrutiny for their claims. We invite them to use a more conservative 
approach. Even if the PSD method employed in \cite{texono} has been 
available for long, all other  experiments have 
consistently chosen to rely on noise abatement on hardware to reduce 
thresholds and increase sensitivity to light WIMPs.
It should be clear from the discussion above that the kind of
exercise attempted by TEXONO requires an extraordinary control of all systematic 
effects and a complete absence of assumptions and bias. Notice we have not 
dwelled on the effect of the ordering of the signal cuts (for which 
contradictory information between text and table can be found in 
\cite{texono}) or choice of integration times. An extensive list 
would be far too long.

In our opinion, the methodology 
followed in \cite{texono} is clearly flawed. The results, as presented, 
untenable. At this point we invite the interested 
reader to develop his or her own opinion.

N.B.: We cannot help but notice in \cite{texono} a reference to 
$\sim$1 kg 
p-type point contact (PPC) detectors \cite{jcap,cogent} like those now 
under consideration by the CoGeNT and {\sc{Majo\-ra\-na}} collaboration, under the 
description of ``a new design in ULEGe''. ``ULEGe'' is the 
commercial brand name for Canberra's line of tiny n-type germanium x-ray 
devices, used and proposed until now by TEXONO, and hardly a ``novel detector''. 
From a technological point of view, ULEGes and PPCs share
little more than a common target material.


\begin{thebibliography}{9}

\bibitem{us} F.T. Avignone {\it et al.}, {\tt arXiv:0806.1341v1}.  
\bibitem{texono} S.T. Lin {\it et al.}, {\tt arXiv:0712.1645v2}.
\bibitem{henry}H. Wong, private communication.
\bibitem{morales}J. Morales {\it et al.}, Nucl. Instr. 
Meth. {\bf A321} 
(1992) 410 (suggested by F.S. Goulding, at the 1st Workshop on Low Background 
Experiments, Berkeley, December, 1989).
\bibitem{jcap} P.S. Barbeau {\it et al.}, JCAP {\bf 09} (2007) 009 
({\tt arXiv:nucl-ex/0701012}).
\bibitem{moralesother} L. Baudis {\it et al.}, Nucl. Instr. 
Meth. {\bf A418} 
(1998) 348; A. Morales {\it et al.}, Phys. Lett. {\bf B532} 
(2002) 8; Astropart. Phys. {\bf 16} 
(2002) 325; P. Benes {\it et al.}, Nucl. Instr. 
Meth. {\bf A569} 
(2006) 737.
\bibitem{cogent} C.E. Aalseth {\it et al.}, {\tt arXiv:0807.0879}.
\bibitem{xray} B.G. Lowe, Nucl. Instr. Meth {\bf A439} 
(2000) 247; M.C. Lepy {\it et al.},  Nucl. Instr. Meth {\bf A439} 
(2000) 239; Nucl. Instr. Meth {\bf A505} 
(2003) 290.
\bibitem{jones}K.W. Jones and H.W. Kraner, Phys. Rev. {\bf A11} 
(1975) 1347.
\end{thebibliography}
\end{document}